%% file: template.tex
\documentclass[cameraready]{Interspeech}

\usepackage{booktabs}
\usepackage{longtable}
\usepackage{pdflscape}
\usepackage{tabularx}
\usepackage{array}
\usepackage{rotating}
\usepackage{placeins}
\usepackage{adjustbox}
\usepackage{dsfont}
\usepackage{subcaption}
\usepackage{float}
\usepackage{multirow}
\usepackage{array}
\usepackage{xurl} 
\usepackage{hyperref}

\title{When Spoof Detectors Travel: Evaluation Across 66 Languages in the Low-Resource Language Spoofing Corpus}

\author[affiliation={1,2}, orcid=0009-0001-8203-1059, correspondingauthor]{Kirill}{Borodin}
\author[affiliation={1,2}, orcid=0009-0001-9935-7514]{Vasiliy}{Kudryavtsev}
\author[affiliation={1}, orcid=0009-0004-7716-8714]{Maxim}{Maslov}
\author[affiliation={1}, orcid=0000-0003-1739-9831]{Mikhail}{Gorodnichev}
\author[affiliation={1}, orcid=0000-0002-5802-5513]{Grach}{Mkrtchian}

\address{
    $^1$ lab260, Moscow Technical University of Communications and Informatics, Russia\\
    $^2$ BitmanagerAI, UAE
}

\email{kborodin.research@gmail.com}


\newcommand{\adfs}{audio deepfakes}

\usepackage{comment}

\begin{document}

\maketitle

\begin{abstract}
We introduce LRLspoof, a large-scale multilingual synthetic-speech corpus for cross-lingual spoof detection, comprising 2,732 hours of audio generated with 24 open-source TTS systems across 66 languages, including 45 low-resource languages under our operational definition. To evaluate robustness without requiring target-domain bona fide speech, we benchmark 11 publicly available countermeasures using threshold transfer: for each model we calibrate an EER operating point on pooled external benchmarks and apply the resulting threshold, reporting spoof rejection rate (SRR). Results show model-dependent cross-lingual disparity, with spoof rejection varying markedly across languages even under controlled conditions, highlighting language as an independent source of domain shift in spoof detection. The dataset is publicly available at https://huggingface.co/datasets/lab260/LRLspoof
and 
https://modelscope.cn/datasets/lab260/LRLspoof.
\end{abstract}

\input{sections/1introduction}
\input{sections/2dataset_description}

\input{sections/3experiments}

\input{sections/4discussion}
\input{sections/5conclusion}

\clearpage
\section{Generative AI Use Disclosure}
This work uses generative models as part of the data creation pipeline: portions of the dataset were synthesized using text-to-speech (TTS) systems to produce spoofed (synthetic) speech samples for anti-spoofing research. Generative AI tools were not used to develop the core scientific contributions beyond this disclosed data generation, and were not used to fabricate or manipulate experimental outcomes, quantitative results, or conclusions. Any optional AI assistance, if used, was limited to language editing (clarity, grammar, and style) and did not introduce new technical content. All authors reviewed and approved the final manuscript and take full responsibility for the dataset, experiments, and claims.
\bibliographystyle{IEEEtran}
\bibliography{mybib}

\end{document}

%% file: sections/1introduction.tex
\section{Introduction}
\label{sec:intro}

Recent progress in text-to-speech (TTS) and voice conversion (VC) has made audio spoofing increasingly practical, raising the stakes for speaker verification and other speech-driven security applications \cite{survey_deepfake}. To keep evaluation reproducible and comparable as attacks evolve, the community has developed shared tasks and benchmarks with standardized protocols and datasets for spoofing countermeasures (CMs), with particular emphasis on generalization to unseen synthesizers and recording conditions \cite{asvspoof2019,asvspoof2021,asvspoof5,add2022,mlaad,indicsynth}. Yet generalization remains a persistent challenge: performance can degrade under domain shifts introduced by new TTS systems and neural vocoders \cite{survey_deepfake,asvspoof2021,librisevoc}, post-processing \cite{asvspoof2021,survey_deepfake}, channel variation \cite{asvspoof2021}, and compression artifacts from conventional and neural codecs \cite{soundstream,encodec}, as well as additive perturbations and adversarial manipulations targeting CMs \cite{rawboost,adv_attacks_asru}.

A comparatively underexplored axis of domain shift is language mismatch. Many widely used public resources are dominated by a small set of high-resource languages\cite{asvspoof2019,asvspoof2021,asvspoof5,add2022,mlaad,indicsynth,survey_deepfake}, which may encourage CMs to rely on language- or phonotactics-correlated artifacts rather than spoofing-specific cues. Recent work reports consistent degradation when CMs trained on one language are evaluated on others, suggesting that language identity can act as a latent bias factor and impede cross-lingual transfer \cite{lang_mismatch,crosslingual_bias}. This matters operationally because spoofing tools are increasingly multilingual, while deployed CMs often face linguistically heterogeneous audio, making cross-lingual evaluation and robustness more important than what is currently specified in many benchmarks.

To address this gap, we introduce LRLspoof, a large-scale multilingual synthetic-speech corpus designed to evaluate spoofing CMs under language mismatch, with particular emphasis on low-resource languages under our operational definition. LRLspoof comprises 2,732 hours of audio spanning 66 languages and 24 open-source TTS synthesizers, enabling controlled evaluation across both (language, synthesizer) axes. We also use LRLspoof as a unified zero-shot diagnostic: for each CM, we set an operating threshold at the Equal Error Rate (EER) on pooled external benchmarks, then apply this fixed threshold without adaptation to each subset to probe cross-lingual generalization under different languages and synthesizers.

\section{Related work}

\input{sections/tables/related_works}

Table~\ref{tab:related_works} compares representative spoofing corpora and benchmarks in terms of multilingual coverage, low-resource languages coverage, number of generation systems, and scale. In this work, we {operationally} define a low-resource language as one with less than 100 hours of {Scripted Speech} in Common Voice 24.0~\cite{commonvoice24}; this criterion is used solely to standardize comparisons within our analyses. Prior benchmarks such as ASVspoof~\cite{asvspoof2019,asvspoof2021,asvspoof5} and ADD~\cite{add2022} have been instrumental for standardizing evaluation, but they are language-centric, making it difficult to perform {controlled} cross-lingual analyses.

More recent multilingual corpora broaden language coverage and/or scale, but they vary in experimental control and intended use. IndicSynth \cite{indicsynth} provides substantial duration yet is restricted to 12 Indic languages and three synthesizers, limiting typological breadth and cross-family generalization studies. AI4T \cite{combei25_interspeech} covers several languages but is comparatively small, while MLADDC \cite{mladdc} is multilingual and large but uses only two vocoder models, making it less tailored to studying language shift under diverse generation conditions. SpeechFake \cite{speechfake} and CVoiceFake \cite{safeear} are multilingual but include fewer low-resource languages under our definition; MLAAD \cite{mlaad} spans many languages and generation systems, but its low-resource coverage is smaller by the same criterion, which can constrain analyses centered on low-resource conditions.

In contrast, our corpus is purpose-built for controlled cross-lingual spoof detection under explicit (language, synthesizer) shifts: we generate spoofed speech using a fixed suite of open-source TTS synthesizers under their upstream implementations and cover a broad set of languages, including many low-resource languages. This design enables direct comparisons where the synthesizer is held fixed and only the language varies, facilitating targeted analyses of language-induced domain shift.


%% file: sections/tables/related_works.tex
\begin{table}[h!]
\centering
\caption{Comparison of representative spoofing corpora. L = number of languages, LRL = number of low-resource languages (per our operational definition). “Models” denotes the number of distinct speech generation systems used to create  audio as reported by each dataset. “Hours” denotes the reported duration of the speech data. Some entries are omitted (--) when the corresponding statistics are not reported.}
\label{tab:related_works}
{
\begin{tabular}{lcccc}
\hline
Dataset & L & LRL & Models & Hours \\
\hline

MLAAD \cite{mlaad}             & 51            & 29  & \textbf{140} & 687.4 \\
IndicSynth \cite{indicsynth}   & 12            & 9  & 3  & 4000 \\

AI4T \cite{combei25_interspeech} & 8           & -- & -- & 13 \\
MLADDC \cite{mladdc}           & 20            & 8 & 2 & 1125 \\



SynHate \cite{synhate}         & 37            & 12  & 1  & 271.7 \\
HateSpeech dataset \cite{mmhate} & 6 & 5 & 1 & -- \\

SpeechFake \cite{speechfake}   & 46            & 21  & 40 & \textbf{4855} \\
CVoiceFake \cite{safeear}      & 5  & 0 & 6 & 215 \\




\textbf{Ours}                  & \textbf{66}   & \textbf{45} & 24 & 2732.17 \\
\hline
\end{tabular}
}
\end{table}

%% file: sections/2dataset_description.tex
\section{Dataset creation}
\label{sec:dataset}

We constructed a multilingual synthetic-speech corpus for spoof detection research. The corpus contains only synthetically generated speech produced with a fixed set of open-source TTS synthesizers across 66 languages.  We include widely used languages alongside many low-resource languages to facilitate controlled cross-lingual generalization experiments.

All synthesized utterances were generated using upstream open-source TTS implementations, executed with no modifications to the codebase. For each (synthesizer, language) condition, we synthesized speech from a predefined set of text prompts using the synthesizer's default configuration for that language (e.g., the default speaker/voice), and we used publicly released pretrained weights when multiple checkpoints were available\footnote{\url{https://huggingface.co/datasets/lab260/LRLspoof}}. We used the following open-source synthesizers: 
\begin{itemize}
    \item \textbf{Classical (non-neural / parametric) TTS:} eSpeak NG~\cite{espeakng}, RHVoice~\cite{rhvoice}, AhoTTS~\cite{ahotts}.
    \item \textbf{Neural supervised TTS (fixed voices):} Silero~\cite{silero}, SpeechT5~\cite{speecht5}, FastPitch~\cite{fastpitch}, Matcha-TTS~\cite{matchatts}, Parler-TTS~\cite{parlertts}, Piper~\cite{pipertts}, MeloTTS~\cite{melotts}.
    \item \textbf{Multilingual / low-resource initiatives:} MMS-TTS~\cite{mms}, Indic-TTS~\cite{indictts}, TurkicTTS~\cite{turkictts}, IMS Toucan~\cite{imstoucan}, QirimtatarTTS~\cite{qirimtatartts}.
    \item \textbf{Generative zero-shot TTS / voice cloning:} XTTS~\cite{xtts}, XTTS2~\cite{xtts2}, OuteTTS~\cite{outetts}, Chatterbox~\cite{chatterbox}, F5-TTS~\cite{f5tts}, CosyVoice~\cite{cosyvoice}, Zonos~\cite{zonos}, Fish-Speech~\cite{fishspeech}, Kokoro~\cite{kokoro}.
\end{itemize}

Text prompts were compiled from multiple publicly available sources to obtain naturally occurring sentence structure and punctuation patterns across languages. Prompt lengths were controlled to yield short utterances suitable for TTS generation and CM evaluation.

\begin{figure*}[t]
    \centering
    \begin{subfigure}[t]{0.49\textwidth}
        \centering
        \includegraphics[width=\linewidth]{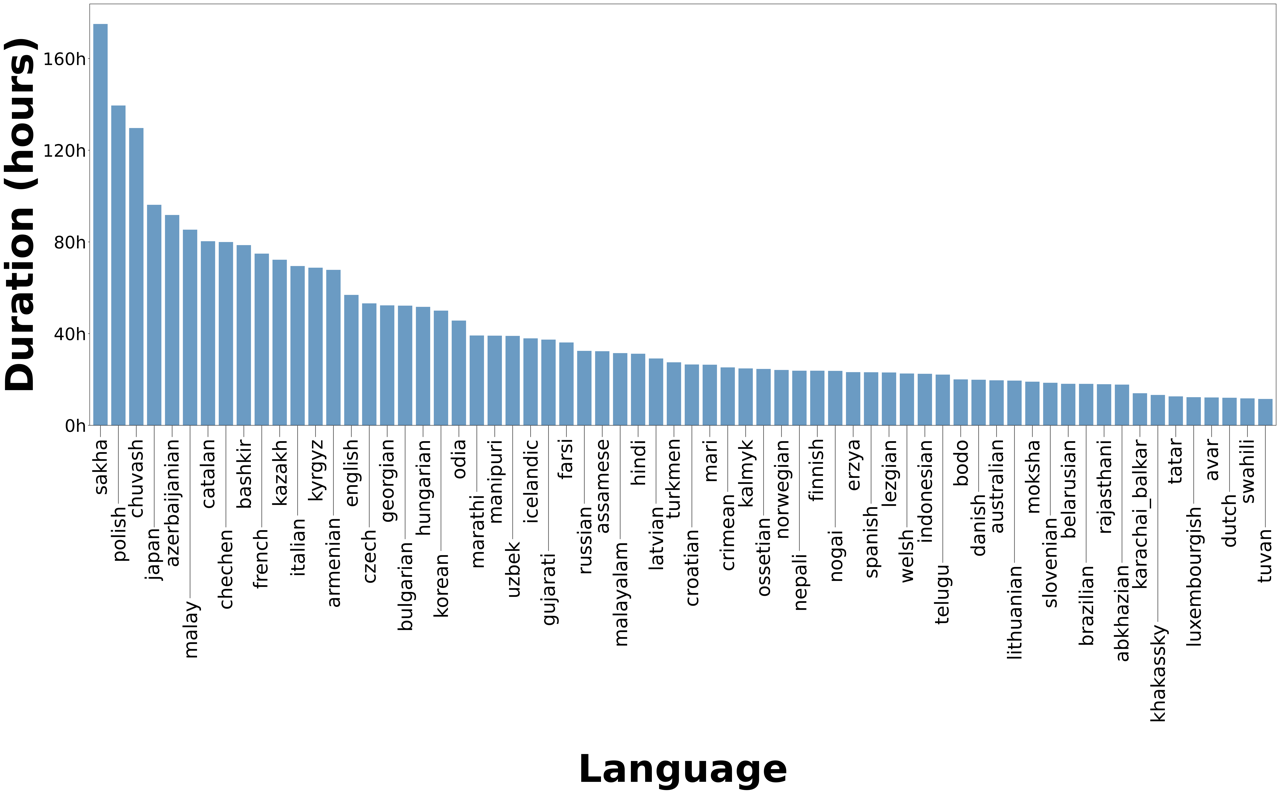}
        \caption{Language distribution.}
        \label{fig:dur_lang}
    \end{subfigure}
    \hfill
    \begin{subfigure}[t]{0.49\textwidth}
        \centering
        \includegraphics[width=\linewidth]{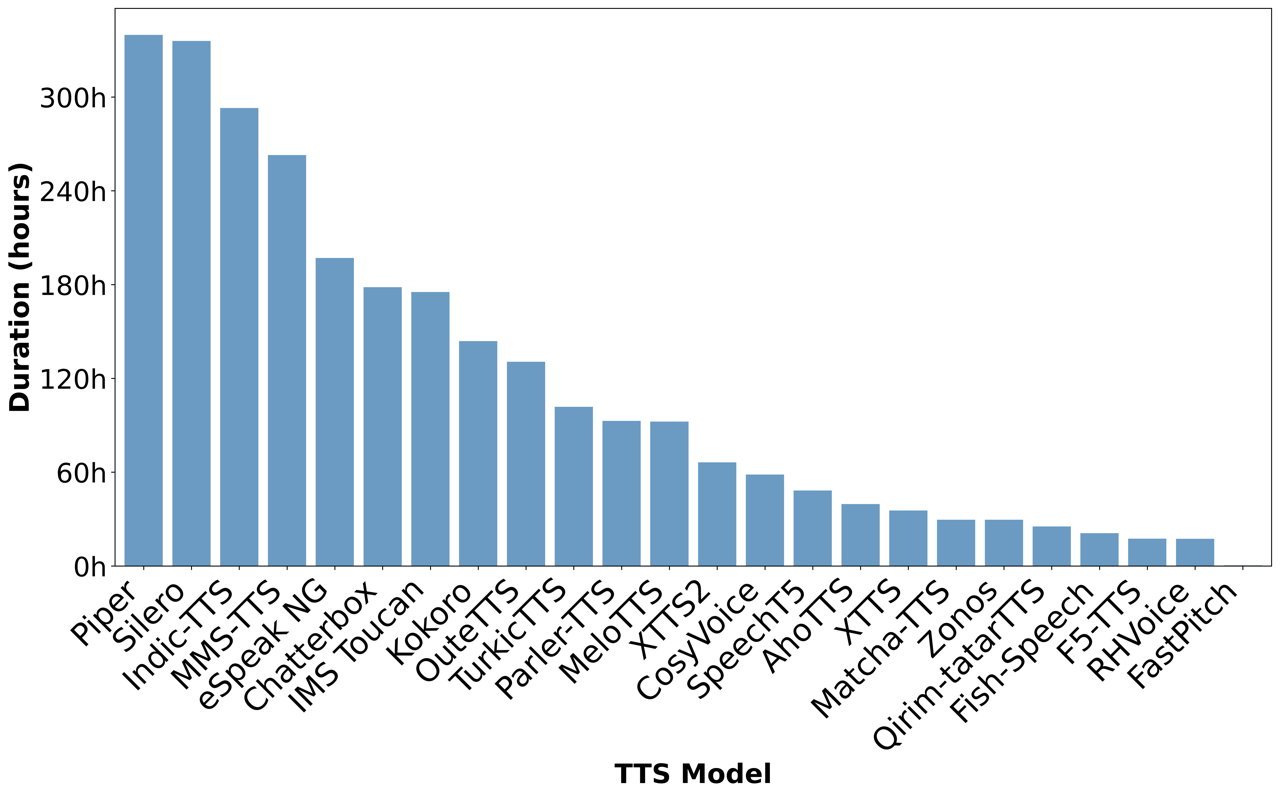}
        \caption{TTS model distribution.}
        \label{fig:dur_tts}
    \end{subfigure}
    \caption{Dataset duration distribution across languages (left) and TTS models (right).}
    \label{fig:duration_distributions}
\end{figure*}

We summarize LRLspoof statistics at three levels. Table~\ref{tab:related_works} situates our corpus relative to prior spoofing and deepfake benchmarks in terms of multilingual and low-resource coverage as well as overall scale. Table~\ref{tab:results_overall} then provides a corpus-level breakdown, reporting per-language duration and the number of synthesizers available for each language, enabling controlled analyses under explicit (language, synthesizer) shifts. Finally, Figure~\ref{fig:duration_distributions} visualizes the resulting duration distributions across languages (Fig.~\ref{fig:dur_lang}) and synthesizers (Fig.~\ref{fig:dur_tts}).

\input{sections/tables/overall_summary}

%% file: sections/tables/overall_summary.tex
\begin{table*}[t]
\centering
\caption{Spoof rejection rate (SRR, \%) across languages. For each language we also report total audio duration (hours) and the number of supported synthesizers (\#Synth).}
\label{tab:results_overall}
\scriptsize
\setlength{\tabcolsep}{2pt}
\begin{adjustbox}{max width=\textwidth}
\begin{tabular}{l S[table-format=4.2] S[table-format=2.0] *{11}{S[table-format=3.2]}  @{\hspace{6pt}}|@{\hspace{6pt}}  l S[table-format=4.2] S[table-format=2.0] *{11}{S[table-format=3.2]}}
\toprule
\multicolumn{3}{c}{} & \multicolumn{11}{c}{\textbf{SRR (\%)}} & \multicolumn{3}{c}{} & \multicolumn{11}{c}{\textbf{SRR (\%)}} \\
\cmidrule(lr){1-3}\cmidrule(lr){4-14}\cmidrule(lr){15-17}\cmidrule(lr){18-28}
\textbf{Lang.} & {\textbf{Hours}} & {\textbf{\#Synth}} & \rotatebox{90}{\textbf{aasist3}} & \rotatebox{90}{\textbf{df\_arena\_1b}} & \rotatebox{90}{\textbf{df\_arena\_500}} & \rotatebox{90}{\textbf{res2tcn}} & \rotatebox{90}{\textbf{rescapsguard}} & \rotatebox{90}{\textbf{sls}} & \rotatebox{90}{\textbf{ssl\_aasist}} & \rotatebox{90}{\textbf{tcm\_add}} & \rotatebox{90}{\textbf{nes2net}} & \rotatebox{90}{\textbf{w2v2\_1b}} & \rotatebox{90}{\textbf{w2v2\_300}} & \textbf{Lang.} & {\textbf{Hours}} & {\textbf{\#Synth}} & \rotatebox{90}{\textbf{aasist3}} & \rotatebox{90}{\textbf{df\_arena\_1b}} & \rotatebox{90}{\textbf{df\_arena\_500}} & \rotatebox{90}{\textbf{res2tcn}} & \rotatebox{90}{\textbf{rescapsguard}} & \rotatebox{90}{\textbf{sls}} & \rotatebox{90}{\textbf{ssl\_aasist}} & \rotatebox{90}{\textbf{tcm\_add}} & \rotatebox{90}{\textbf{nes2net}} & \rotatebox{90}{\textbf{w2v2\_1b}} & \rotatebox{90}{\textbf{w2v2\_300}} \\
\midrule
sakha & 175.07 & 3 & 99.45 & 76.86 & 63.60 & 67.54 & 42.68 & 42.05 & 44.61 & 40.67 & 11.58 & 56.86 & 64.33 & polish & 139.51 & 5 & 83.87 & 43.42 & 45.67 & 42.65 & 52.24 & 12.23 & 27.12 & 22.76 & 7.95 & 21.90 & 30.78 \\
chuvash & 129.68 & 5 & 99.64 & 74.41 & 40.43 & 31.58 & 37.54 & 44.47 & 76.61 & 69.97 & 32.85 & 50.52 & 79.38 & japanese & 96.18 & 4 & 95.00 & 80.71 & 40.58 & 0.54 & 35.38 & 0.10 & 1.18 & 1.12 & 0.10 & 3.46 & 89.05 \\
azerbaijani & 91.76 & 4 & 98.64 & 60.04 & 32.38 & 57.76 & 35.26 & 47.41 & 55.54 & 58.42 & 44.79 & 39.50 & 71.24 & malay & 85.34 & 1 & 99.96 & 69.76 & 64.69 & 28.10 & 47.28 & 7.34 & 18.23 & 11.25 & 1.39 & 0.96 & 63.26 \\
catalan & 80.31 & 2 & 92.14 & 16.42 & 66.32 & 26.24 & 65.27 & 18.84 & 31.30 & 15.32 & 0.01 & 9.14 & 71.68 & chechen & 79.96 & 1 & 99.86 & 94.95 & 57.10 & 95.44 & 2.83 & 97.24 & 99.66 & 99.94 & 99.80 & 16.25 & 99.91 \\
bashkir & 78.61 & 4 & 99.06 & 47.75 & 32.34 & 62.65 & 31.50 & 45.82 & 51.96 & 57.24 & 45.76 & 35.02 & 71.87 & french & 74.88 & 2 & 77.90 & 90.79 & 38.15 & 0.28 & 6.91 & 5.14 & 18.80 & 18.48 & 6.32 & 19.33 & 75.50 \\
kazakh & 72.23 & 3 & 95.05 & 41.91 & 33.24 & 72.51 & 62.16 & 62.24 & 66.34 & 68.28 & 39.45 & 52.23 & 41.91 & italian & 69.47 & 2 & 99.88 & 95.14 & 71.42 & 4.18 & 34.35 & 15.68 & 8.89 & 7.92 & 0.04 & 1.28 & 99.99 \\
kyrgyz & 68.76 & 3 & 99.99 & 82.05 & 69.38 & 68.78 & 91.73 & 75.64 & 77.12 & 77.26 & 81.72 & 87.39 & 100.00 & armenian & 67.82 & 3 & 99.80 & 74.49 & 33.41 & 66.45 & 40.28 & 37.82 & 44.95 & 47.11 & 46.10 & 46.72 & 76.12 \\
english & 56.89 & 4 & 93.33 & 87.11 & 40.00 & 0.81 & 13.34 & 11.31 & 10.56 & 12.89 & 5.78 & 7.91 & 62.27 & czech & 53.18 & 3 & 79.95 & 24.11 & 72.27 & 12.08 & 64.38 & 51.18 & 54.37 & 50.28 & 42.09 & 26.06 & 67.42 \\
georgian & 52.31 & 3 & 67.25 & 69.54 & 55.95 & 33.37 & 36.93 & 35.88 & 56.94 & 60.06 & 36.16 & 20.35 & 91.68 & bulgarian & 52.19 & 2 & 97.70 & 82.14 & 33.49 & 91.30 & 55.52 & 98.10 & 99.34 & 99.22 & 95.70 & 20.38 & 99.50 \\
hungarian & 51.68 & 2 & 95.40 & 66.38 & 71.80 & 19.30 & 80.19 & 88.38 & 92.92 & 87.58 & 63.21 & 43.50 & 99.06 & korean & 50.02 & 2 & 51.74 & 85.68 & 98.50 & 0.00 & 37.71 & 3.58 & 2.08 & 3.46 & 1.84 & 0.72 & 75.16 \\
odia & 45.69 & 1 & 100.00 & 65.90 & 81.79 & 71.06 & 84.67 & 30.66 & 68.14 & 55.54 & 40.92 & 38.14 & 99.83 & marathi & 39.17 & 1 & 99.47 & 91.82 & 94.83 & 31.51 & 81.84 & 43.22 & 56.56 & 66.26 & 43.91 & 1.43 & 99.77 \\
manipuri & 39.11 & 1 & 91.35 & 97.79 & 54.88 & 2.99 & 14.77 & 98.07 & 90.21 & 99.99 & 95.28 & 59.15 & 99.99 & uzbek & 38.99 & 2 & 96.52 & 58.36 & 63.95 & 33.88 & 25.74 & 45.72 & 62.00 & 61.30 & 78.80 & 63.27 & 63.26 \\
icelandic & 37.93 & 1 & 100.00 & 25.17 & 47.23 & 54.43 & 41.42 & 0.31 & 0.74 & 2.28 & 0.00 & 0.02 & 25.21 & gujarati & 37.38 & 1 & 98.27 & 97.15 & 17.66 & 15.73 & 5.83 & 34.83 & 91.12 & 82.32 & 6.54 & 55.41 & 99.99 \\
farsi & 36.14 & 1 & 99.97 & 31.98 & 95.82 & 16.37 & 95.57 & 20.44 & 28.62 & 16.45 & 0.01 & 0.55 & 64.23 & russian & 32.48 & 2 & 54.95 & 99.50 & 93.18 & 25.67 & 48.06 & 49.26 & 46.88 & 46.49 & 13.74 & 26.86 & 99.23 \\
assamese & 32.32 & 1 & 33.62 & 100.00 & 94.52 & 0.02 & 4.76 & 6.33 & 37.18 & 28.40 & 97.78 & 0.00 & 99.08 & malayalam & 31.50 & 1 & 100.00 & 54.31 & 92.95 & 0.60 & 54.60 & 37.30 & 14.46 & 50.65 & 0.95 & 1.80 & 99.51 \\
hindi & 31.21 & 2 & 98.18 & 93.66 & 6.59 & 0.00 & 0.48 & 0.00 & 18.85 & 7.71 & 0.09 & 1.23 & 97.06 & latvian & 29.15 & 1 & 91.94 & 71.69 & 57.76 & 12.46 & 2.98 & 2.28 & 9.14 & 11.61 & 0.17 & 0.24 & 46.19 \\
turkmen & 27.47 & 1 & 100.00 & 100.00 & 26.00 & 100.00 & 86.11 & 95.31 & 96.48 & 99.93 & 100.00 & 85.72 & 100.00 & croatian & 26.54 & 1 & 92.42 & 95.40 & 98.77 & 0.11 & 14.95 & 81.90 & 91.02 & 90.84 & 99.99 & 87.29 & 91.46 \\
mari & 26.43 & 1 & 100.00 & 99.50 & 88.31 & 99.23 & 3.25 & 79.70 & 99.61 & 99.99 & 95.59 & 66.56 & 100.00 & crimean\_tatar & 25.27 & 1 & 99.79 & 93.41 & 99.82 & 52.13 & 49.51 & 89.58 & 64.62 & 75.90 & 25.02 & 42.11 & 98.52 \\
kalmyk & 24.80 & 1 & 100.00 & 99.99 & 73.29 & 92.10 & 0.64 & 69.81 & 98.58 & 99.90 & 99.48 & 63.73 & 99.99 & ossetian & 24.60 & 1 & 99.99 & 98.24 & 36.66 & 35.12 & 6.36 & 85.51 & 97.74 & 99.99 & 90.21 & 19.91 & 99.78 \\
norwegian & 24.15 & 1 & 99.49 & 82.68 & 77.74 & 13.47 & 52.68 & 3.50 & 8.54 & 6.79 & 1.84 & 0.20 & 48.02 & nepali & 23.83 & 1 & 99.92 & 46.38 & 4.69 & 80.87 & 25.16 & 8.05 & 29.53 & 26.90 & 0.07 & 52.31 & 9.07 \\
finnish & 23.83 & 1 & 91.57 & 63.41 & 71.97 & 27.17 & 50.18 & 11.04 & 23.98 & 13.11 & 1.28 & 0.45 & 17.02 & nogai & 23.74 & 1 & 100.00 & 99.88 & 49.00 & 98.06 & 2.86 & 58.84 & 99.28 & 99.98 & 97.92 & 8.22 & 100.00 \\
erzya & 23.20 & 1 & 100.00 & 95.04 & 99.69 & 81.44 & 86.52 & 94.63 & 95.97 & 88.22 & 3.92 & 81.03 & 99.93 & spanish & 23.14 & 1 & 99.83 & 31.64 & 99.99 & 26.28 & 99.98 & 1.36 & 0.61 & 2.96 & 0.00 & 0.04 & 100.00 \\
lezgian & 23.05 & 1 & 96.00 & 44.27 & 88.53 & 0.00 & 24.76 & 19.44 & 60.76 & 57.01 & 51.33 & 26.16 & 100.00 & welsh & 22.59 & 1 & 0.24 & 29.46 & 12.91 & 13.10 & 92.00 & 83.72 & 91.71 & 92.74 & 97.32 & 12.30 & 99.99 \\
indonesian & 22.46 & 1 & 99.98 & 3.98 & 99.98 & 2.84 & 99.82 & 0.11 & 0.03 & 0.33 & 0.00 & 0.00 & 95.82 & telugu & 22.12 & 1 & 73.68 & 76.38 & 8.45 & 3.28 & 0.56 & 0.01 & 0.55 & 0.73 & 0.00 & 0.02 & 100.00 \\
bodo & 20.03 & 1 & 55.02 & 39.43 & 64.70 & 9.97 & 36.67 & 17.24 & 13.17 & 31.98 & 12.33 & 1.78 & 99.86 & danish & 19.87 & 1 & 99.85 & 7.01 & 87.29 & 12.55 & 54.95 & 2.94 & 5.41 & 7.35 & 0.14 & 0.00 & 93.26 \\
australian\_english & 19.62 & 1 & 99.95 & 74.82 & 98.42 & 0.04 & 69.08 & 0.18 & 0.48 & 2.13 & 0.02 & 0.00 & 99.87 & lithuanian & 19.51 & 1 & 98.60 & 98.51 & 93.42 & 0.21 & 7.76 & 3.54 & 3.07 & 2.80 & 13.06 & 3.01 & 99.41 \\
moksha & 19.03 & 1 & 98.89 & 86.13 & 93.56 & 11.89 & 94.93 & 67.57 & 66.81 & 69.80 & 5.00 & 56.70 & 99.86 & slovenian & 18.55 & 1 & 100.00 & 17.11 & 88.67 & 85.49 & 85.10 & 69.00 & 59.34 & 51.55 & 1.38 & 48.14 & 97.07 \\
belarusian & 18.10 & 1 & 98.09 & 98.64 & 21.47 & 28.46 & 1.56 & 3.19 & 31.86 & 20.35 & 48.19 & 38.26 & 63.61 & brazilian\_portuguese & 18.08 & 1 & 100.00 & 69.53 & 2.97 & 9.81 & 0.65 & 0.12 & 23.71 & 10.94 & 0.00 & 0.89 & 99.99 \\
rajasthani & 17.93 & 1 & 40.94 & 99.89 & 89.20 & 0.09 & 2.57 & 3.16 & 16.69 & 30.41 & 33.96 & 1.25 & 90.35 & abkhazian & 17.76 & 1 & 99.62 & 59.24 & 54.02 & 0.00 & 2.98 & 10.82 & 42.19 & 61.29 & 53.03 & 13.30 & 99.94 \\
karachai\_balkar & 14.00 & 1 & 98.02 & 80.44 & 90.87 & 0.00 & 6.75 & 13.69 & 40.52 & 57.12 & 81.69 & 4.88 & 99.99 & khakassky & 13.25 & 1 & 88.46 & 95.26 & 99.94 & 61.48 & 96.34 & 16.37 & 19.34 & 14.96 & 2.17 & 85.92 & 2.42 \\
tatar & 12.63 & 1 & 97.52 & 42.36 & 34.97 & 61.74 & 23.24 & 74.14 & 79.64 & 86.40 & 31.73 & 81.26 & 33.34 & luxembourgish & 12.28 & 1 & 93.83 & 96.36 & 99.53 & 31.60 & 64.99 & 25.82 & 24.47 & 17.30 & 0.19 & 3.86 & 93.62 \\
avar & 12.13 & 1 & 100.00 & 84.02 & 100.00 & 65.97 & 84.70 & 100.00 & 98.51 & 100.00 & 71.00 & 69.41 & 100.00 & dutch & 12.01 & 1 & 58.24 & 66.94 & 18.44 & 0.22 & 6.84 & 5.06 & 21.29 & 17.84 & 2.29 & 2.38 & 30.04 \\
swahili & 11.75 & 1 & 99.62 & 70.12 & 58.81 & 31.69 & 49.88 & 5.46 & 9.74 & 9.02 & 2.53 & 0.22 & 28.09 & tuvan & 11.50 & 1 & 97.01 & 94.29 & 95.14 & 0.00 & 6.46 & 8.69 & 33.58 & 46.63 & 95.65 & 2.94 & 99.98 \\

\bottomrule
\end{tabular}
\end{adjustbox}
\end{table*}

%% file: sections/3experiments.tex
\section{Experimental setup}
\label{sec:experiments}



\subsection{Spoofing countermeasures}
\label{sec:cms}
We evaluate 11 publicly available spoofing CMs spanning classical spectro-temporal architectures and large self-supervised encoders: \texttt{aasist3}~\cite{aasist3}, \texttt{df\_arena\_1b}~\cite{df_arena}, \texttt{df\_arena\_500}~\cite{df_arena}, \texttt{res2tcn}~\cite{borodin2024capsule}, \texttt{rescapsguard}~\cite{borodin2024capsule}, \texttt{sls}~\cite{sls}, \texttt{ssl\_aasist}~\cite{tak2022automatic}, \texttt{tcm\_add}~\cite{truong24_tcmadd}, \texttt{nes2net}~\cite{nes2net}, \texttt{w2v2\_1b}~\cite{w2v2_1b}, and \texttt{w2v2\_300}~\cite{w2v2_300}.

\subsection{Spoof-only evaluation at a transferred EER operating point}
\label{sec:srr}
Let $s(x)$ denote the CM score for utterance $x$; higher scores indicate bona fide.
Given a threshold $\tau$, the spoof rejection rate (SRR) on our spoof-only corpus is
\[
\mathrm{SRR}(\tau)=\frac{1}{N}\sum_{i=1}^{N}\mathds{1}\big[s(x_i)<\tau\big],
\]
where $\{x_i\}_{i=1}^{N}$ are spoofed utterances.
Since our corpus contains no bona fide speech, we cannot compute an EER directly on it; instead, we evaluate SRR at a {transferred} operating point $\tau_{\mathrm{EER}}$ calibrated on pooled external benchmarks (Section~\ref{sec:threshold_calib}). SRR is spoof-only: a high value is desirable but does not constitute a full security operating point, since FRR on the target domain is not measured (Section~\ref{sec:implications_limitations}).

\subsection{Threshold calibration on external benchmarks}
\label{sec:threshold_calib}
We calibrate thresholds on pooled external benchmarks that contain both bona fide and spoofed speech: ASVspoof5 (ASV5)~\cite{asvspoof5}, ASVSpoof2021 LA (ASLA) and DF (ASDF)~\cite{asvspoof2021}, in-the-wild (InTW)~\cite{in_the_wild}, DFADD (DFDD)~\cite{dfadd}, and ADD2022 (ADD2)~\cite{add2022}.
For each CM, we concatenate all trials across these datasets, compute the EER on the resulting pool, and set $\tau_{\mathrm{EER}}$ at the point where false-accept and false-reject rates are equal.  
This protocol probes robustness under explicit language and synthesizer shift at a fixed operating point, while avoiding the need for bona fide speech in the target corpus. Table~\ref{tab:eer} reports the resulting per-benchmark and pooled EER values used for calibration. We deliberately use one transferred threshold per CM rather than per-language thresholds, which would be a supervised oracle requiring labeled target-language data — precisely the supervision our zero-shot setting avoids.

\input{sections/tables/eers}

\subsection{Controlled evaluation with fixed model and synthesizer}
\label{sec:controlled_pairs}

\input{sections/tables/robust_table}

To isolate the role of language, we perform controlled comparisons under the same fixed, transferred threshold, where we fix (i) the CM and (ii) the TTS system, and vary only the language.
If performance changes substantially under these constraints, this provides evidence that  language-induced distribution shift is an independent driver of robustness degradation, beyond synthesizer identity alone.
Table~\ref{tab:popular_vs_lowresource_pairs} reports illustrative controlled contrasts at the transferred EER operating point.

%% file: sections/tables/eers.tex
\begin{table}[!t]
\centering
\caption{Equal error rate (EER, \%) of the evaluated spoofing countermeasures on each external benchmark used for threshold calibration}
\label{tab:eer}
\scriptsize
\setlength{\tabcolsep}{4pt}
\begin{adjustbox}{max width=\columnwidth}
\begin{tabular}{l c c c c c c c}
\toprule
\textbf{Model} & ASV5 & ASLA & ASDF & InTW & DFDD & ADD2 & Pooled \\
\midrule
\texttt{aasist3}        & 41.001 & 37.407 & 33.099 & 39.626 & 17.973 & 47.192 & 41.094 \\
\texttt{df\_arena\_1b}   & 35.333 & 40.137 & 42.994 & 17.598 & 16.545 & 42.139 & 37.003 \\
\texttt{df\_arena\_500}  & 34.992 & 24.725 & 29.021 & 30.400 & 10.216 & 41.835 &  33.443 \\
\texttt{res2tcn}        & 37.620 & 19.130 &  19.883 & 49.246 & 34.136 &  49.538 &  38.798\\
\texttt{rescapsguard}   & 49.748 & 18.768 &  28.478 & 49.017 & 33.538 &41.316 & 45.085\\
\texttt{sls}            & 17.668 & 7.714 & 4.220 & 7.453 & 23.837 & 33.951 & 16.346 \\
\texttt{ssl\_aasist}     & 20.004 & {7.506} & \textbf{3.254} & 8.313 & 12.292 & 42.482 & 17.402  \\
\texttt{tcm\_add}        & 19.505 & \textbf{6.655} & 3.444 & 7.767 & 9.833 &  35.252 & 5.983\\
\texttt{nes2net}        & \textbf{7.465} & 18.334 & 10.725 & 23.93 &  7.442 & \textbf{30.327} & 16.485 \\
\texttt{w2v2\_1b}         & 26.508 & 8.66 & 6.389 & \textbf{4.288} &  \textbf{6.512} & 40.933 &  24.935\\
\texttt{w2v2\_300}        & 39.375 & 19.483 & 11.553 & 4.35 & 10.648 & 42.482 & 32.985\\
\bottomrule
\end{tabular}
\end{adjustbox}
\end{table}

%% file: sections/tables/robust_table.tex
\begin{table*}[t]
\centering
\scriptsize
\setlength{\tabcolsep}{2pt}
\renewcommand{\arraystretch}{1.15}

\caption{Most illustrative controlled low-resource language contrasts under a fixed TTS system. Diff (pp) denotes the absolute SRR difference in percentage points between the popular language and the low-resource language at the same EER-calibrated operating point.}
\label{tab:popular_vs_lowresource_pairs}

\sisetup{
  round-mode = places,
  round-precision = 2,
  table-number-alignment = center
}
\newcommand{\na}{\multicolumn{1}{c}{--}}

\begin{adjustbox}{max width=\textwidth}
\begin{tabular}{l l *{11}{S[table-format=2.2]}}
\toprule
\multicolumn{2}{c}{} & \multicolumn{11}{c}{\textbf{Diff (pp) by CM}} \\
\cmidrule(lr){1-2}\cmidrule(lr){3-13}
\textbf{Language pair} & \textbf{TTS} &
\rotatebox{90}{\textbf{aasist3}} &
\rotatebox{90}{\textbf{df\_arena\_1b}} &
\rotatebox{90}{\textbf{df\_arena\_500}} &
\rotatebox{90}{\textbf{res2tcn}} &
\rotatebox{90}{\textbf{rescapsguard}} &
\rotatebox{90}{\textbf{sls}} &
\rotatebox{90}{\textbf{ssl\_aasist}} &
\rotatebox{90}{\textbf{tcm\_add}} &
\rotatebox{90}{\textbf{nes2net}} &
\rotatebox{90}{\textbf{w2v2\_1b}} &
\rotatebox{90}{\textbf{w2v2\_300}} \\
\midrule

English--Polish  & \texttt{Parler-TTS} &
\na & \na & 82.06 & \na & \na & \na & \na & \na & 61.60 & \na & 94.46 \\

Hindi--Odia  & \texttt{Indic-TTS} &
\na & \na & \na & 59.22 & 75.38 & \na & \na & \na & \na & \na & \na \\

Dutch--Bashkir  & \texttt{MMS-TTS} &
\na & \na & \na & 77.50 & \na & \na & \na & \na & \na & \na & \na \\

Hindi--Manipuri  & \texttt{Indic-TTS} &
\na & \na & \na & \na & \na & \na & 57.52 & 87.88 & \na & 63.90 & \na \\

Luxembourgish--Georgian  & \texttt{Piper} &
92.27 & 77.65 & 64.69 & \na & \na & \na & 56.86 & 62.34 & \na & \na & \na \\

Ossetian--Sakha  & \texttt{Silero} &
\na & \na & \na & 53.96 & 86.39 & \na & \na & \na & 56.44 & 58.20 & 50.88 \\

Assamese--Gujarati  & \texttt{Indic-TTS} &
\na & \na & 76.85 & \na & \na & \na & 59.74 & 52.42 & \na & 71.86 & \na \\

Moksha--Chechen  & \texttt{Silero} &
\na & \na & \na & 88.01 & \na & 50.06 & 58.92 & 51.09 & \na & \na & \na \\

Danish--Nepali  & \texttt{Piper} &
\na & \na & 82.61 & 65.08 & \na & \na & \na & \na & \na & 55.40 & 77.67 \\

Indonesian--Nepali  & \texttt{Piper} &
\na & \na & 95.27 & 70.69 & \na & \na & \na & \na & \na & 55.40 & 68.14 \\
  
Danish--Welsh  & \texttt{Piper} &
99.51 & \na & 74.33 & \na & \na & \na & 69.39 & 80.19 & \na & \na & \na \\

Indonesian--Welsh  & \texttt{Piper} &
99.59 & \na & 86.99 & \na & \na & \na & 72.32 & 84.39 & \na & \na & \na \\

\bottomrule
\end{tabular}
\end{adjustbox}
\end{table*}

%% file: sections/4discussion.tex
\section{Results and discussion}
\label{sec:discussion}

\input{sections/tables/4metric_table}

\subsection{Overall robustness under threshold transfer}
\label{sec:overall_results}
We first summarize overall spoof rejection performance when transferring EER-calibrated thresholds from pooled external benchmarks to our corpus. Table~\ref{tab:results_overall} shows that threshold transfer can yield widely varying spoof rejection rates (SRR) across both languages and CMs: for example, \texttt{aasist3} achieves strong SRR on several languages (e.g., 93.33\% on English and 99.86\% on Chechen), while \texttt{w2v2\_300} degrades sharply on others (e.g., 62.27\% on English and 30.78\% on Polish).

To provide a compact view beyond the per-language breakdown, Table~\ref{tab:srr_summary} aggregates SRR in two ways: (i) \emph{mean} across languages (each language has equal weight) and (ii) \emph{pooled} across all utterances (languages are weighted by their number of samples). We report both aggregations for the full set of languages and for the low-resource subset, highlighting that several models degrade noticeably under low-resource conditions, revealing robustness gaps that can be obscured by broader language coverage.

\subsection{Cross-lingual robustness and disparity analysis}
\label{sec:crosslingual_results}

A key observation is that cross-lingual disparity is often model-dependent: for instance, \texttt{nes2net} fails almost completely for some languages (e.g., 0.01\% on Catalan) but rejects nearly all spoofs for others (e.g., 99.80\% on Chechen and 97.32\% on Welsh in Table~\ref{tab:results_overall}). Similarly, \texttt{df\_arena\_500} is strong on Marathi (94.83\%) but collapses on Nepali (4.69\%). These gaps suggest that CMs can inadvertently rely on language artifacts, leading to fragile behavior under language shift, even when the attack generator remains fixed.

\subsection{Synthesizer effects, interactions, and controlled pairs}
\label{sec:synth_results}
We analyze SRR across language–synthesizer subsets to quantify synthesizer effects and their interaction with language. Table~\ref{tab:results_overall} already indicates that some models are sensitive to particular language subsets that are tied to specific synthesizer coverage, and Table~\ref{tab:popular_vs_lowresource_pairs} makes this interaction explicit under controlled conditions.

In particular, even when the synthesizer is held fixed, SRR can shift by tens of percentage points across languages for the same CM. For example, under \texttt{Parler-TTS}, the English--Polish contrast reaches 94.46 pp for \texttt{w2v2\_300} and 82.06 pp for \texttt{df\_arena\_500}; under \texttt{Piper}, the Danish--Welsh contrast reaches 99.51 pp for \texttt{aasist3}. These results highlight that language mismatch and synthesizer choice can interact: a given TTS system can be reliably rejected in one language while bypassing the same detector in another.

Using the controlled comparisons in Table~\ref{tab:popular_vs_lowresource_pairs}, we further highlight representative cases where robustness differences persist even when both the CM and synthesizer are held fixed. For instance, with \texttt{Indic-TTS}, Hindi--Odia shows a 59.22 pp gap for \texttt{res2tcn} and a 75.38 pp gap for \texttt{rescapsguard}, despite identical generation conditions. 
These controlled contrasts provide direct evidence that language identity can act as an independent source of domain shift in spoof detection.

\subsection{Implications and limitations}
\label{sec:implications_limitations}
These results suggest that cross-lingual robustness is a major failure mode for current anti-spoofing models: the same CM can behave very differently across languages, even when the attack generator is held fixed. Accordingly, we position LRLspoof as a diagnostic corpus for analyzing cross-lingual failure modes, and we emphasize disparity patterns across language pairs and CM families over any single pooled SRR number. 
This motivates training/evaluation that targets language shift, reporting cross-lingual disparity rather than only language-matched results, and model designs that avoid language-specific artifacts.


Our evaluation is deliberately spoof-only. We do not include target-language bona fide speech because simply adding bona fide from existing datasets would introduce a domain confound: a countermeasure could learn to discriminate domain A vs. domain B rather than bona fide vs. spoof. A sound target-language benchmark would instead require matched bona fide recordings collected under comparable language, speaker, text, and channel conditions.  We  position LRLspoof as a spoof-side cross-lingual stress test for existing CMs in a zero-shot setting. This design inherits the dependence on external calibration benchmarks (sec.~\ref{sec:threshold_calib}), which may not span all deployment languages and channels.

%% file: sections/tables/4metric_table.tex
\begin{table}[t]
\centering
\caption{Summary SRR (\%) aggregated across languages. \texttt{MSRR (All)}: mean SRR across all languages; \texttt{PSRR (All)}: pooled SRR across all languages; \texttt{MSRR (Low)}: mean SRR across low-resource languages; \texttt{PSRR (Low)}: pooled SRR across low-resource languages.}
\label{tab:srr_summary}
\scriptsize
\setlength{\tabcolsep}{4pt}
\begin{adjustbox}{max width=\columnwidth}
\begin{tabular}{l S[table-format=2.2] S[table-format=2.2] S[table-format=2.2] S[table-format=2.2]}
\toprule
\textbf{Model} & {\textbf{MSRR (All)}} & {\textbf{PSRR (All)}} & {\textbf{MSRR (Low)}} & {\textbf{PSRR (Low)}} \\
\midrule
\texttt{aasist3}        & 90.40 & 91.40 & 93.28 & 94.28 \\
\texttt{df\_arena\_1b}   & 71.07 & 70.53 & 73.62 & 74.89 \\
\texttt{df\_arena\_500}  & 63.47 & 58.50 & 67.31 & 62.47 \\
\texttt{res2tcn}        & 33.07 & 35.37 & 37.38 & 42.94 \\
\texttt{rescapsguard}   & 41.29 & 40.56 & 40.81 & 40.65 \\
\texttt{sls}            & 36.04 & 34.85 & 39.66 & 39.97 \\
\texttt{ssl\_aasist}     & 45.62 & 43.48 & 50.09 & 49.34 \\
\texttt{tcm\_add}        & 45.93 & 43.20 & 51.30 & 49.66 \\
\texttt{nes2net}        & 34.23 & 30.85 & 38.34 & 35.61 \\
\texttt{w2v2\_1b}         & 26.79 & 26.07 & 30.28 & 30.14 \\
\texttt{w2v2\_300}        & 80.51 & 78.32 & 83.90 & 81.29 \\
\bottomrule
\end{tabular}
\end{adjustbox}
\end{table}

%% file: sections/5conclusion.tex
\section{Conclusion}
\label{sec:conclusion}

Across 11 public CMs evaluated on LRLspoof at a transferred EER threshold, spoof rejection varies with specific language–synthesizer pairs rather than the synthesizer alone. This interaction effect indicates that many CMs do not transfer reliably under language shift, and merits explicit consideration in future evaluation and model design.